\documentclass[prb,floatfix,twocolumn,amsmath,amssymb,showpacs]{revtex4}
\usepackage{graphicx}

\begin{document}

\title{Quantum dot with ferromagnetic leads: a densiti-matrix renormalization 
group study}

\author{C. J. Gazza, M. E. Torio, and J. A. Riera}
\affiliation{
Instituto de F\'{\i}sica Rosario, Consejo Nacional de
Investigaciones Cient\'{\i}ficas y T\'ecnicas,\\
Universidad Nacional de Rosario, Rosario, Argentina}

\date{\today}

\begin{abstract}
A quantum dot coupled to ferromagnetically polarized
one-dimensional leads is studied numerically using the density-matrix 
renormalization group method. Several real space properties and 
the local density of states at the dot are computed.
It is shown that this local density of states is suppressed by the
parallel polarization of the leads. In this case we are able to 
estimate the length of the Kondo cloud, and to relate its behavior 
to that suppression. Another important result of our study is that the 
tunnel magnetoresistance as a function of the quantum dot on-site 
energy is minimum and negative at the symmetric point.
\end{abstract}


\maketitle
%%%%%%%%%%%%%%%%%%% introduction %%%%%%%%%%%%%%%%%%%%%%%%%%%%%%%%%%

The relentless pursuit of smaller, faster, more efficient
electronic devices has led to an increasing interest in
mesoscopic and nanoscopic devices behaving as quantum dots (QD),
using semiconductor technology\cite{goldhaber} and nanotubes as
components\cite{tsukagoshi} among other possibilities.
Some of these devices allow the exploitation of the electron spin
in addition to its charge with various potential applications in
the so-called spintronics.\cite{wolf} In particular, a spintronic
device connecting a QD to ferromagnetic leads has already been 
experimentally studied.\cite{pasupathy}

These nanoscopic devices have also been growing in complexity.
For example a device consisting of two QDs has been developed to
show a nonlocal control of one QD over the other via a RKKY
interaction.\cite{craig} Another example is the device where an
appropriate arrangement of electrodes defines a section of a nanotube 
as the QD.\cite{biercuk} This QD turns out to be then connected to essentially
one-dimensional leads. In the first example, a real space description
of the magnetic correlations would be important to understand the
internal working of this device. In the second example, the effects
of finite length of the nanotube sections connected to the QD would
have to be considered.

Among many techniques used to study these kinds of problems, which
have been developed in the context of the Kondo effect,\cite{hewson}
the numerical renormalization group (NRG) has provided
many important and essentially exact results. However, this technique
cannot tackle the increasingly complex devices such as the above
mentioned ones. For this reason, we propose in this paper the application
of a real space numerical technique, the density-matrix renormalization
group (DMRG)\cite{white}, which has been extensively used to study
quasi-one-dimensional  strongly correlated electron systems.\cite{review}
This technique is ideally suited to provide detailed real space
information, such as spin-spin correlations or electron site occupancies,
required to understand nanoscopic devices. In addition, it
works on finite size systems. This is not a disadvantage when bulk leads
are involved in the device because in most cases one could resort to some
kind of extrapolation. On the contrary, it can provide results for finite
systems such as the example mentioned before.

In this paper, we will study the simplest device in spintronics,
the spin valve. This device consists of a single-level
QD attached to two noninteracting leads. On these leads, a
ferromagnetic polarization is introduced by an applied magnetic
field, or in other words, by a rigid displacement of spin-up and
spin-down electron bands. The central quantities to study are the
conductance when the polarization of the leads are parallel
($G_P$) and antiparallel ($G_{AP}$). The main measure of the performance 
of the spin valve is the tunnel magnetoresistance (TMR) defined as
$TMR=(G_P - G_{AP})/G_{AP}$. This device has been studied before
using different approximations\cite{martinek0,bdong,weynman}.
These analytical techniques are not exact and in fact they lead to
some degree of controversy around central issues such as the
behavior of the conductance when the leads have a parallel
polarization. Some of these controversial issues have been partially
settled using NRG.\cite{martinek,sanchez}

%%%%%%%%%%%% model %%%%%%%%%%%%%%%%%%

As in previous studies on this system, the Hamiltonian model for a
QD located at the center of the chain is a generalization of the
Anderson model defined as
\begin{eqnarray}
{\cal H} = &-& t \sum_{i=\leq -2,\sigma} (c^{\dagger}_{i \sigma}
c_{i+1 \sigma} + H.c. ) - h_L \sum_{i=\leq -1} S^z_i  \nonumber \\
&-& t \sum_{i=\geq 1,\sigma} (c^{\dagger}_{i \sigma} c_{i+1
\sigma}+ H.c. ) - h_R \sum_{i=\geq 1} S^z_i  \nonumber \\
&-& t' \sum_{\sigma} (c^{\dagger}_{-1 \sigma} c_{0 \sigma} +
c^{\dagger}_{0 \sigma} c_{1 \sigma} + H.c. ) \nonumber\\
&+& \epsilon ~ n_{0} + U  n_{0, \uparrow} n_{0, \downarrow}
\label{hamilt}
\end{eqnarray}
\noindent where conventional notation was used. 
We adopt $t$ as the scale of energy. 
The magnetic field on the leads can be set in two configurations,
$h_L = h_R= h$, which corresponds to a parallel (P) polarization
of the leads, and $h_L =-h_R= h$ which corresponds to the
antiparallel (AP) one. In the following, $h > 0$
favors a positive polarization.
Notice that the leads are described  by a real-space tight-binding model
which is in principle more realistic than the ones implicit in the NRG
treatment.
%%%%%%%%%%%%%%%%% results %%%%%%%%%%%%%%%%%%%%%%%%%%%%%%%%%%%%%%%

Model equation(\ref{hamilt}) was studied by the DMRG on $L=63,79$ and 95 
chains with open boundary conditions. Most of the results
shown below were obtained for two sets of parameters: $U=1$,
$t'=0.4$ and $U=4$, $t'=0.8$. For both sets of parameters the
effective Kondo coupling at the symmetric point is $J=4t'^2/U=0.64$,
although strictly speaking this relationship is only valid for $U
>> t'$. There are, however, many important differences between both 
sets of parameters as we will show below.
%%%%%%%%%%%%%%%%% Figure 1 %%%%%%%%%%%%%%%%%%%%%%%%%%%%%%%%%%
%\vspace{2 mm}
\begin{figure}%[ht]
\includegraphics[width=0.40\textwidth]{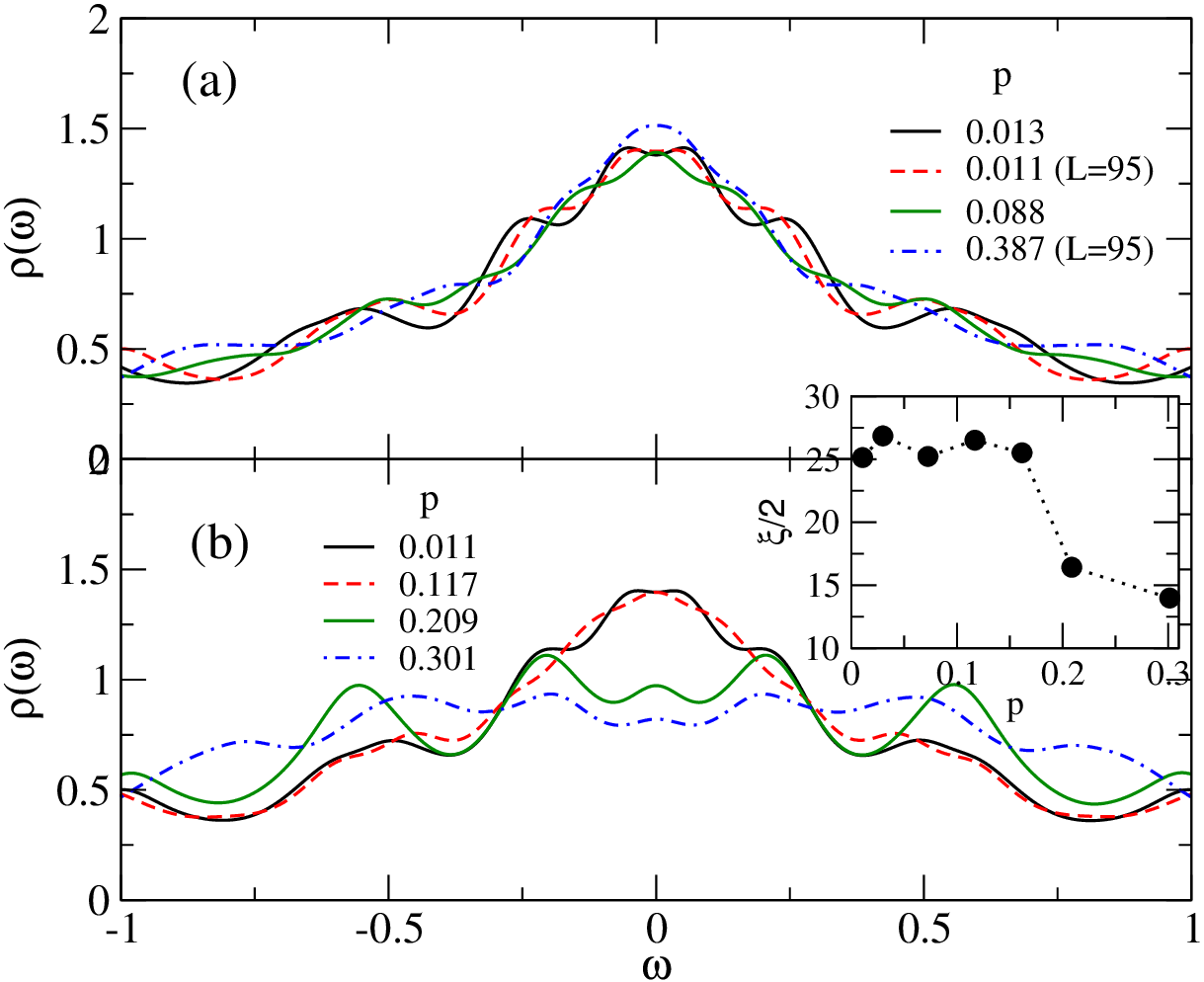}
\caption{(Color online) The LDOS for $U=1$, $t'=0.4$, and
$\epsilon=-0.5$, for various values of the polarization $p$, (a)
AP magnetization for the $L=79$ chain, and (b) P, $L=95$. 
The inset shows the length of the ``compensation cloud" (see text) 
for the P case, $L=95$.} \label{fig1}
\vspace{-4 mm}
\end{figure}
%%%%%%%%%%%%%%%%%%%%%%%%%%%%%%%%%%%%%%%%%%%%%%%%%%%%%%%%%%%%%%

The main quantity that we have studied is the local density of
states (LDOS), $\rho(\omega)$, at the QD. In the first place, from
this quantity it is possible to evaluate the conductance in the
linear response regime. In the second place, recent advances in
scanning tunneling microscopy (STM) have made it possible to
directly measure this quantity. Since in our DMRG calculation the
ground state vector is measured when the two added sites are at
the center  of the chain (symmetrical configuration), the QD then
is one of these two sites which are exactly treated. Then, we
adopt the approximation of applying the creation and annihilation
operators  at the QD on the ground state vector and then determine
$\rho(\omega)$ following the well-known continued fraction
formalism. A more accurate approach would be, after the
application of each of those creation and annihilation operators,
to run additional sweeps for an enlarged density
matrix.\cite{hallberg} In any case, the truncation of the Hilbert 
space is the essential source of error in the DMRG, and to estimate the
precision of our approach we have compared results for various
numbers of retained states $M$, from $M=500$ to 1000. We would
also like to stress the fact that the conductance in linear
response is related to the LDOS near $\omega=0$
where the approximation is more precise.\cite{note1} In addition, 
we have computed standard properties such as the electron occupancy 
of each site, $<n_{i,\sigma}>$ ($\sigma=\uparrow, \downarrow$) and
spin-spin correlations from the QD, 
$S(j)=<S_0^z~S_j^z>-<S_0^z><S_j^z>$.

All the results presented correspond to the half-filled system.
For the AP case, we work in the $S^z=1/2$
sector. For the parallel polarization, we work in the $S^z$
subspace which minimizes the total energy, and hence it depends on
$h$. The polarization of each lead $p_{\alpha}$ ($\alpha = L, R$)
is defined as $p_{\alpha}=(n_{\alpha \uparrow} - n_{\alpha
\downarrow})/(n_{\alpha \uparrow} + n_{\alpha \downarrow})$. Then,
the polarization $p$ of the P (AP) configuration is $(p_L + p_R)/2$
[$(p_L - p_R)/2$]. For small clusters, and specially for the $U=1$ case, 
it is difficult to get $p_L=p_R$, i.e. the same numbers of solitons on 
each lead. In these cases the AP case shows spurious suppression of the
LDOS.\cite{torio}
%%%%%%%%%%%%%%%%% Figure 2 %%%%%%%%%%%%%%%%%%%%%%%%%%%%%%%%%%%%%%%
%\vspace{3 mm}
\begin{figure}%[hb]
\includegraphics[width=0.40\textwidth]{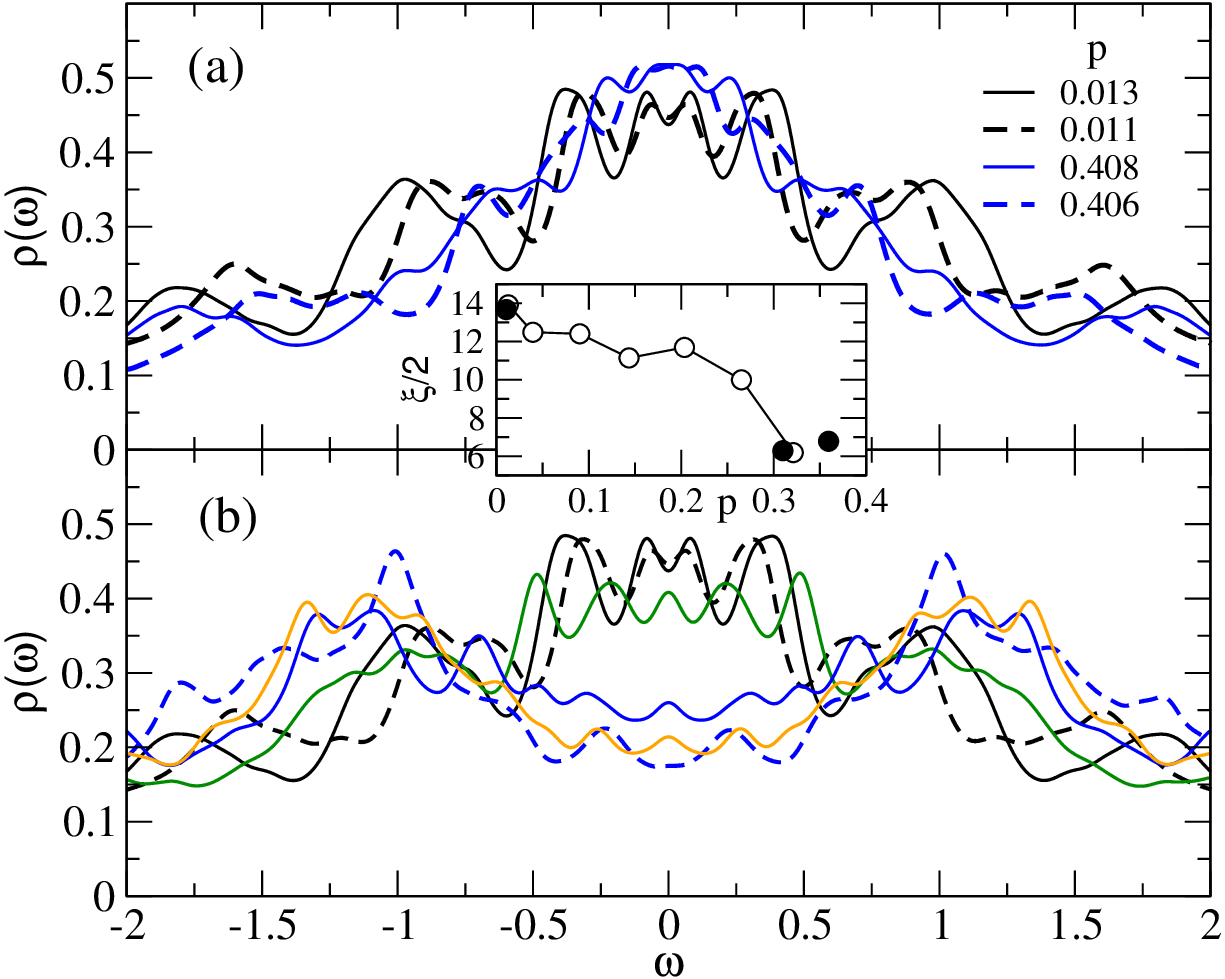}
\caption{(Color online) The LDOS for $U=4$, $t'=0.8$, and $\epsilon=-2.0$, for various
values of the polarization $p$, (a) antiparallel, $L=79$ (solid
lines), $L=95$ (dashed lines); (b) parallel configuration, from
top to bottom at $\omega=0$: $L=79$, $p=0.013, 0.203, 0.321,
0.426$ (solid lines); $L=95$, $p=0.011, 0.399$ (dashed lines).The
inset shows the length of the ``compensation cloud" (see text) for
the P case, $L=79$ (open circles), $L=95$ (filled circles).}
\label{fig2}
\vspace{-4 mm}
\end{figure}
%%%%%%%%%%%%%%%%%%%%%%%%%%%%%%%%%%%%%%%%%%%%%%%%%%%%%%%%%%%%%%%%%%%

In Fig.~\ref{fig1} we show the LDOS at the QD, 
$U=1$, $t'=0.4$, at the symmetric point $\epsilon = -U/2$, for 
several values of the polarization $p$. In this
figure and all the following similar ones, we adopted a Lorentzian
broadening of the peaks of $\delta=0.1$ In the first place, for
$p=0.013$, corresponding to $h=0$, the LDOS is the one expected in
the Kondo problem. The small splitting of the Kondo peak is a
finite size effect. For the case of an AP configuration
(Fig.~\ref{fig1}(a)), a small value of $p=0.088$ leads to a slight
modification of the LDOS, that virtually remains unmodified by
further increasing the polarization of the leads, at least in the
range examined ($p \leq 0.3$). The largest value considered, $p =
0.3$, is obtained for $h=2.2$ ($L=95$). On the other hand, for the P
configuration (Fig.~\ref{fig1}(b)) it can be seen as a suppression of
the Kondo peak for the polarization $p \geq 0.2$. For small values of
$p$ the peak at $\omega=0$ remains roughly unmodified. As we will
show below, this different dependence of the LDOS with $p$ for the
AP and P arrangements leads to the expected behavior of a spin valve. 
The inset in Fig. 1 shows, for the P magnetization,  the length of the 
``compensation cloud"\cite{gubernatis}, a possible measure of the Kondo 
cloud, defined in such a way that the sum of the correlations 
$S(j)$ ($-\xi /2 \leq j \leq \xi/2, j \neq 0$) is equal to 
$0.9 S(0)$.\cite{torio} The reduction of $\xi$ with $p$, due to the
better screening of the minority spin at the QD by the majority
spins on the leads, is consistent with the
suppression of the Kondo resonance observed in the LDOS.

Qualitatively the same behavior can be observed for the case of
the parameters $U=4$, $t'=0.8$, at the symmetric point
$\epsilon=-2.0$. In this case, the Kondo
resonance has a smaller weight as compared with the previous
set of parameters. In Fig.~\ref{fig2}(a), which shows the LDOS
around $\omega=0$, it can be seen again that for the AP case the
LDOS is not much changed as $p$ is increased. In contrast, for the
P case, the Kondo peak is strongly suppressed in the presence of
polarization. This suppression is considerably stronger than for
the previous set of parameters, $U=1$, $t'=0.4$. We have also
examined the size dependence of these results by computing the
LDOS on the $L=95$ chain. Although the position of the peaks, as
expected, moves to lower frequencies (in absolute value), the
overall behavior with $p$ is similar to that for $L=79$. As in the
previous case, the length of the ``compensation cloud" (shown in
the inset in Fig. 2) for the P polarization decreases with increasing $p$
in agreement with the suppression of the Kondo resonance.
It is interesting to note that, consistently with this behavior, the
value of 
$\langle (S^z_0)^2\rangle$ also decreases from $\sim 0.18$ for 
$p\sim 0$  to $\sim 0.10$ for $p\sim 0.4$. Notice that in the
full Kondo regime $\langle (S^z_0)^2\rangle \sim 0.25$.

%%%%%%%%%%%%%%%%% Figure 3 %%%%%%%%%%%%%%%%%%%%%%%%%%%%%%%%%%%%%%
%\vspace{10 mm}
\begin{figure}%[ht]
\includegraphics[width=0.40\textwidth]{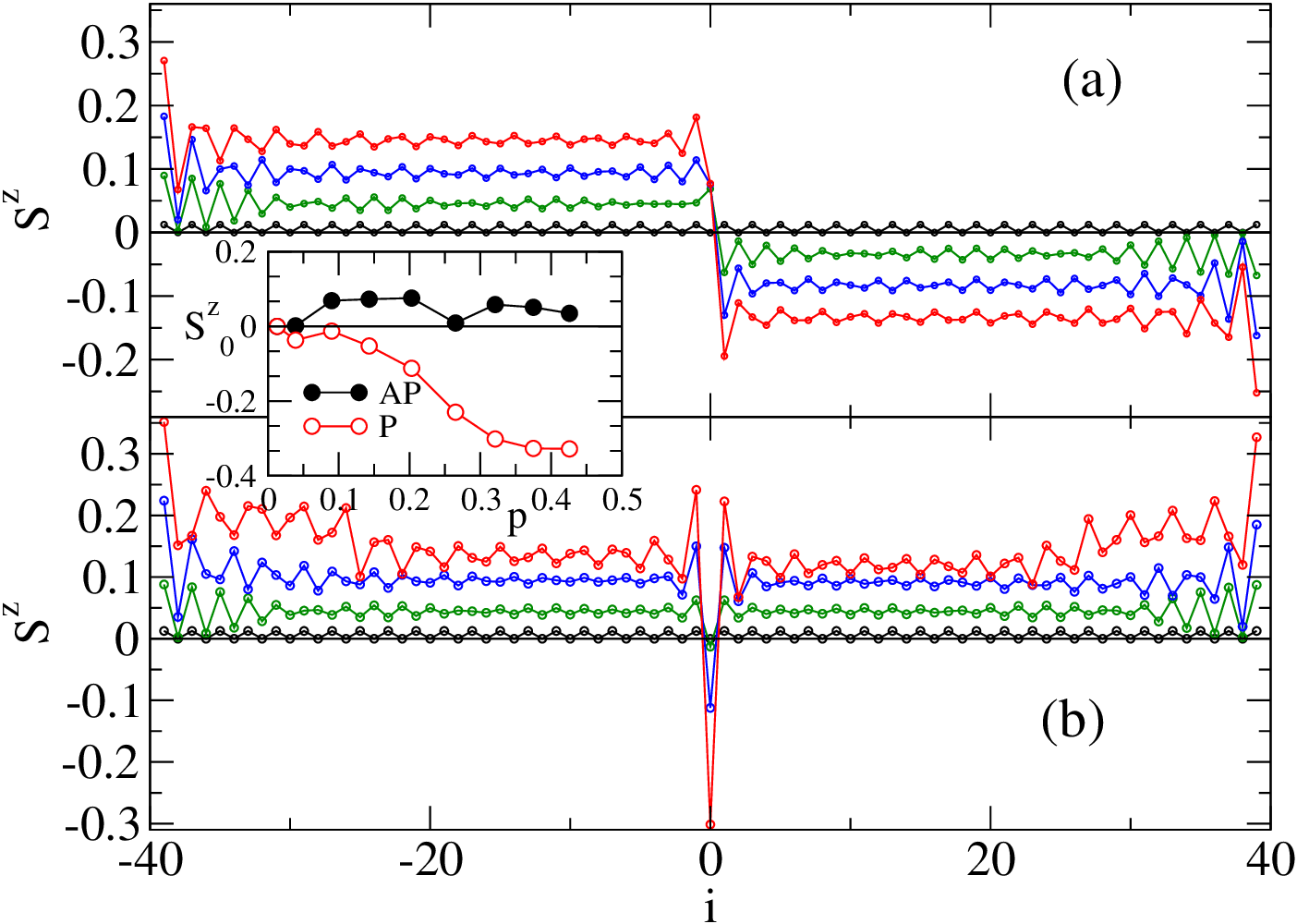}
\caption{(Color online) $S^z_i$ for the  $L=79$ chain, $U=4$, $t'=0.8$,
$\epsilon=-2.0$, for $p\approx 0.01, 0.1, 0.2$ and 0.3, from
bottom to top on the left lead; (a) antiparallel, (b) parallel
configuration. The inset shows $S^z$ at the QD as a function of
the polarization}
\label{fig3}
\vspace{-3 mm}
\end{figure}
%%%%%%%%%%%%%%%%%%%%%%%%%%%%%%%%%%%%%%%%%%%%%%%%%%%%%%%%%%%%%%%%%

The mean value of $S^z_i$ is shown in Fig.~\ref{fig3} for $U=4$,
$t'=0.8$, and $\epsilon=-2.0$, for the AP and P polarizations in 
panels (a) and (b), respectively.
The inhomogeneous way in which the
sites are polarized, due to the open boundary conditions of our
clusters, is clearly visible. We have also observed this behavior 
for the same parameters on the 95-site cluster, suggesting that it
is not a mere finite size effect.
The different behavior of the AP and P configurations can be
traced to the behavior of the mean value of the spin at the QD.
The apparently different behavior
of $S_0^z=S^z_{QD}$ as a function of the polarization is shown in
the inset. For the AP arrangement $S_0^z$ is almost not affected by
$p$, while for the P arrangement $S_0^z$ goes to the minimum
possible value by increasing $p$. This behavior of $S_0^z$ in the 
P case is well known and it has been shown that a magnetic field 
applied to the QD could in fact reduce the suppression of the LDOS 
discussed above.\cite{martinek,weynman}
%%%%%%%%%%%%%%%%% Figure 4 %%%%%%%%%%%%%%%%%%%%%%%%%%%%%%%%%%%%%%%
\begin{figure}%[ht]
\includegraphics[width=0.3\textwidth]{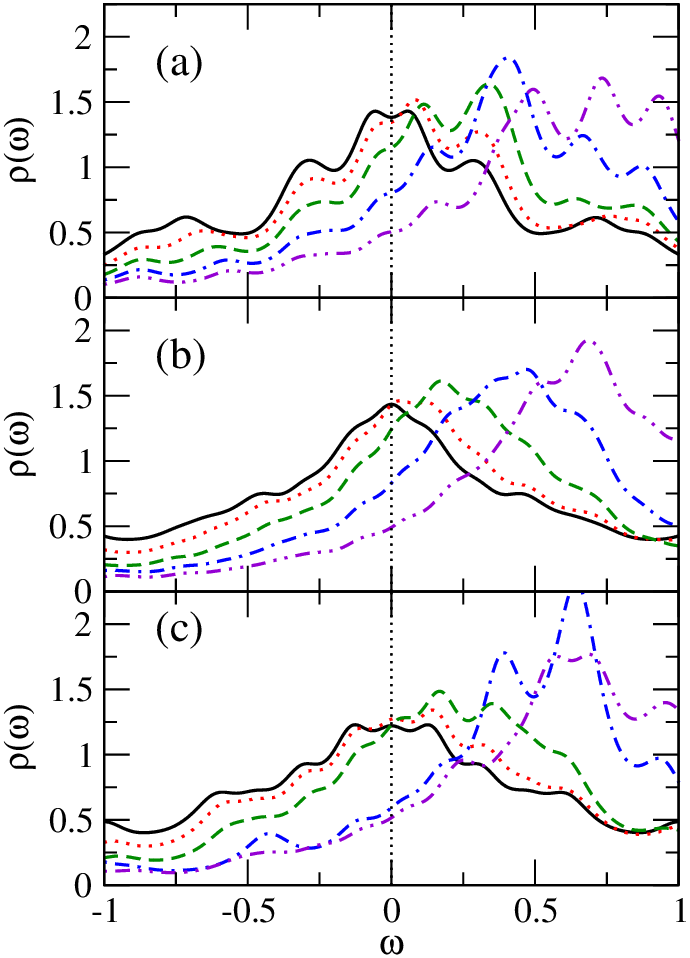}
\caption{(Color online) The LDOS for the  $L=79$ chain: $U=1$, $t'=0.4$, and
$\epsilon=$-0.5 (solid line), -0.25 (dotted line), 0.0 (dashed
line), 0.25 (dotted-dashed line), 0.5 (dotted-dotted-dashed line) for (a)
$p=0.0$, (b) $p=0.30$, AP configuration, and (c) $p=0.30$, parallel
configuration.}
\label{fig4}
\vspace{-4 mm}
\end{figure}
%%%%%%%%%%%%%%%%%%%%%%%%%%%%%%%%%%%%%%%%%%%%%%%%%%%%%%%%%%%%%%%%%

\begin{figure}%[ht]
\includegraphics[width=0.4\textwidth]{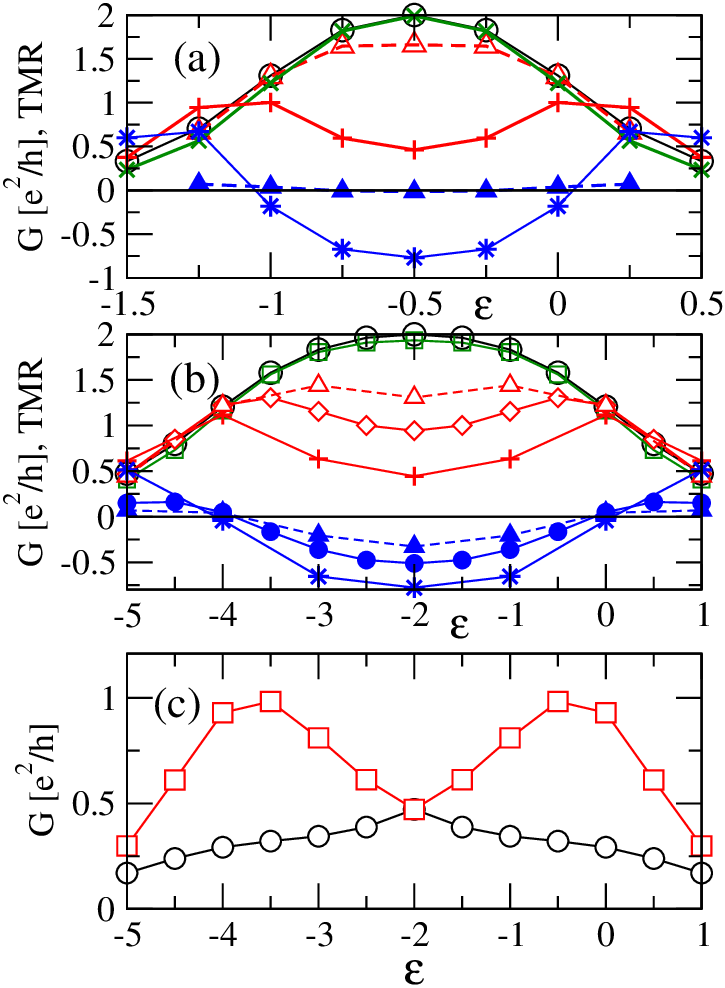}
\caption{(Color online) (a) Conductance as a function of the gate voltage for
$U=1$, $t'=0.4$, $L=79$, and $p=0$ (open circles). Results for $L=63$, 
$p=0.30$: P configuration (open triangles), TMR (filled triangles), and
$L=95$, $p=0.35$ AP (crosses), P (plusses), and TMR (stars). (b) Same 
as (a) for $U=4$, $t'=0.8$. Symbols for $L=63$ and $L=95$ same as (a).
Results for $L=79$: AP (squares), P (diamonds), and TMR (filled circles). 
The values of the polarization are
$p=0.31$ ($L=63$), $p=0.32$ ($L=79$), and $p=0.36$ ($L=95$). 
(c) Up-spin (circles) and down-spin (squares) contributions to
$G_P$ for the parameters of (b), $L=79$.}
\label{fig5}
\vspace{-3 mm}
\end{figure}

We now study the most important issue in this kind of systems,
i.e., the behavior of the conductance and hence of the TMR as a
function of the applied voltage. To understand this behavior we
show in Fig.~\ref{fig4} the LDOS on the $L=79$ chain for the
parameters $U=1$, $t'=0.4$, lead polarization $p=0.3$, and various
values of $\epsilon$. In Fig.~\ref{fig4}(a) we show for comparison
the unpolarized case. The relevant quantity is the LDOS at
$\omega=0$ which is proportional to the conductance, $G=(2 \pi
t'^2/t)\rho(\omega=0)$, in units of $e^2/h$. $G$ as a function of $\epsilon$ 
has the typical form of a dome as it can be seen in Fig.~\ref{fig5}(a).
However, $\rho(\omega=0)$ for a finite cluster is an artifact
of the Lorentzian broadening $\delta$ of the peaks adopted in our
calculations. Although $\rho(\omega=0)$ could be computed in this
way\cite{note2}, we prefer to discuss the results of
Fig.~\ref{fig4} at a qualitative level. The LDOS for $p=0.3$ in
the AP polarization is shown Fig.~\ref{fig4}(b). Its $\omega$
dependence for various values of $\epsilon$ is very similar to the
one for the unpolarized case. On the other hand, for the P
configuration, the LDOS, depicted in Fig.~\ref{fig4}(c), shows at
$\omega=0$ a nonmonotonic behavior as a function of the gate
voltage. This nonmonotonic behavior, when translated to the
conductance, is crucial to understand the $\epsilon$ dependence of
the TMR.
%%%%%%%%%%%%%%%%% Figure 5 %%%%%%%%%%%%%%%%%%%%%%%%%%%%%%%%%%%%%%%
%%%%%%%%%%%%%%%%%%%%%%%%%%%%%%%%%%%%%%%%%%%%%%%%%%%%%%%%%%%%%%%%%%

Our most relevant results are shown in
Fig.~\ref{fig5}. To compute the conductance we use the Friedel sum 
rule\cite{hewson}, $G_\sigma=\sin^2{\pi n_\sigma}$, where
$n_\sigma$ is the occupancy at the QD of an electron with spin
$\sigma$. The Friedel sum rule is valid for arbitrary polarization in the P 
case, except for negligible higher order corrections\cite{martinek}. 
The total conductance is $G=G_\uparrow + G_\downarrow$. 

Figure~\ref{fig5}(a) shows the conductance for the unpolarized case, 
AP and P polarizations, for $U=1$, $t'=0.4$, and Fig.~\ref{fig5}(b) 
for $U=4$, $t'=0.8$. The conductance
for the AP configuration is virtually indistinguishable from the
unpolarized one for $L=79$. On the other hand, $G_P$ has a minimum
at the symmetric point, then increases as $\epsilon$ moves apart
from that point, and finally it merges with the curve for the
unpolarized case, decreasing for large values of the gate voltage.
Then, in the most important region for application purposes, i.e.
near the symmetric point, the TMR is negative with a minimum
precisely at this point. This result is in contradiction
with the one reported in Ref.~\onlinecite{sanchez}, showing a positive 
value at the symmetric point. This difference may be traced to the model 
we adopted for the leads, which implies loosing the particle-hole 
symmetry for the P case, which is present in
Ref.~\onlinecite{sanchez}.\cite{martinek1} 
Notice that for the case of Fig.~\ref{fig5}(b), there 
is a clear trend as the chain size is increased, that is, a deepening 
of $G_P$ around the symmetric point, with a subsequent decrease in the 
TMR. Notice that the lead polarization is slightly different in the three 
clusters. This trend is not that clear for the case of Fig.~\ref{fig5}(a), 
since here the smaller value of $U$ requires larger lattice sizes.

Finally, the up- and down-spin contributions to $G_P$ are shown in
Fig.~\ref{fig5}(c) for the couplings of Fig.~\ref{fig5}(b). 
At the symmetric point $G_\uparrow = G_\downarrow$. As $\epsilon$ 
is moved away from the symmetric point, the conductance of the 
minority spin becomes dominating, reaching perfect conductance, at
$\epsilon \sim -U$ and $\epsilon \sim 0$, and it accounts for the 
increase of the total conductance.

%%%%%%%%%% conclusions %%%%%%%%%%%%%%%%%%%%%%%%%%%%%%%%%%%%%%%%%%%%%%%

In summary, we have shown the possibility of applying the DMRG technique to 
the problem of transport through quantum dots.
For the particular problem of a quantum dot coupled to polarized leads 
we have shown that the LDOS at the QD is suppressed by the 
P polarization of the leads. In addition, we have shown that the
TMR as a function of the quantum dot on-site
energy is minimum and negative at the symmetric point, a result
which is at variance with those reported in previous studies. This is a central
issue in these kinds of devices. In addition to this, we have shown that 
this technique allowed us to estimate the length of the Kondo cloud, and 
to relate its behavior with the suppression of the Kondo peak, correlated in 
turn with the conductance. This emphasizes the fact that real space properties, 
which are accessible by the DMRG, and are very difficult to obtain by alternative 
techniques, are very important to understand the behavior of 
these devices. 

%A further development would be to implement the ``embedding"
%procedure to improve the description of bulk leads.\cite{embedding}

\acknowledgments We acknowledge useful discussions with A. E. Feiguin, S. Maekawa, 
J. Martinek, and G. B. Martins. This work was supported by grant No.
PICT 03-12409 (ANPCYT).


\begin{thebibliography}{}
\bibitem{note1} Our calculation starts to be less precise as we
move away from $\omega=0$ since then higher excited states are
involved.
\end{thebibliography}
\end{document}